\documentclass[a4paper,11pt]{article}
\usepackage{pos}

\title{Phenomenological developments for event-by-event fluctuations of conserved charges}
\ShortTitle{Phenomenological developments for e-by-e fluctuations of conserved charges}

\author*{Volodymyr Vovchenko}

\affiliation{Nuclear Science Division, Lawrence Berkeley National Laboratory,\\
   1 Cyclotron Road, 94720 Berkeley, CA, USA}

\emailAdd{vovchenko@lbl.gov}

\abstract{A brief overview of the recent developments concerning theoretical description of event-by-event fluctuations in heavy-ion collisions is presented, with an emphasis on the role of exact conservation laws and calculations based on relativistic hydrodynamics.}

\FullConference{%
	The International conference on Critical Point and Onset of Deconfinement - CPOD2021\\
  	15 – 19 March 2021\\
 	Online - zoom
 }

\begin{document}
\maketitle

\section{Introduction}

One of the most interesting open questions in high-energy physics is the determination of QCD phase structure at finite temperatures and baryon densities.
While the first-principle lattice QCD simulations have established a smooth QCD chiral crossover transition at vanishing baryochemical potential~\cite{Aoki:2006we}, such calculations are not directly available at finite $\mu_B$ due to the sign problem.
Indirect lattice methods like Taylor expansion around $\mu_B/T = 0$ or analytic continuation from imaginary $\mu_B/T$ suggest that the crossover nature of the QCD transition persists at small $\mu_B/T \lesssim 2-3$~\cite{Bazavov:2017dus,Vovchenko:2017gkg,Borsanyi:2021sxv}, disfavoring the presence of the QCD critical point in that region of the phase diagram.
Complementary to the first-principle theory calculations, heavy-ion collisions provide experimental access to the QCD phase diagram.
In contrast to lattice QCD, however, a broader range of baryon densities is accessible in the experiment by varying the collision energy, including those inaccessible by the current lattice methods.

The analysis of hadron multiplicities in the framework of statistical hadronization model indicates that a high degree of chemical equilibration is reached in heavy-ion collisions, and allows one to map the chemical freeze-out stage to the QCD phase diagram~(see e.g. an overview in~\cite{Andronic:2017pug}).
The mean multiplicities alone, however, cannot directly reveal the finer details of QCD matter properties.
Instead, event-by-event fluctuations of particle numbers have been suggested as sensitive probes of the QCD phase structure~\cite{Stephanov:1998dy,Stephanov:1999zu,Jeon:2000wg,Asakawa:2000wh}.
To understand the basic theoretical motivation behind the use of event-by-event fluctuations one can consider the structure of the grand canonical partition function in statistical mechanics.
Assuming a system with a single conserved number $N$ for simplicity, it reads
\begin{equation}
    \ln Z^{\rm gce} (T,V,\mu) = \ln \left[ \sum_N e^{\mu N/T} \, Z^{\rm ce} (T,V,N) \right].
\end{equation}
Here $Z^{\rm ce} (T,V,N) \propto P(N)$ is the canonical partition function which is proportional to the probability distribution of $N$.
Up to the irrelevant normalization factor, $\ln Z^{\rm gce} (T,V,\mu)$ is proportional to the cumulant generating function $G_N(t)$ for $N$:
\begin{equation}
G_N(t) \equiv \langle e^{tN} \rangle = \sum_{N} e^{tN} P(N) = \sum_{n=1}^{\infty} \kappa_n \frac{t^n}{n!}
\end{equation}
with $t =\mu/T$.
This implies
\begin{equation}
\kappa_n \propto \frac{\partial^n (\ln Z^{\rm gce})}{\partial \mu^n},
\end{equation}
i.e. the cumulants $\kappa_n$ measure the chemical potential derivatives of the equation of state.
This observation forms the basis for different uses of fluctuations in heavy-ion collisions, including the search for the QCD critical point with high-order cumulants~\cite{Hatta:2003wn,Stephanov:2008qz}, experimental test of lattice QCD and the chiral crossover transition at $\mu_B = 0$~\cite{Friman:2011pf}, as well as complementing the mean multiplicities in extracting the freeze-out properties and testing the equilibrium hypothesis on the level of fluctuations~\cite{Bazavov:2012vg,Borsanyi:2014ewa}.

Various event-by-event fluctuation measurements have now been performed in a broad collision energy range by different experiments, including the ALICE Collaboration at LHC~\cite{ALICE:2019nbs}, the STAR Collaboration within RHIC beam energy scan~(RHIC-BES) programme~\cite{STAR:2020tga,STAR:2021iop}, the NA61/SHINE Collaboration at SPS~\cite{Gazdzicki:2017zrq}, and the HADES Collaboration at GSI~\cite{HADES:2020wpc}.

\section{Theory vs experiment: caveats}

\begin{itemize}
    \item Accuracy of the grand-canonical ensemble 
    
    The basic theoretical argument for fluctuations is based on grand-canonical statistical mechanics, where the system can freely exchange conserved charges with the heat bath.
    Incidentally, this is also the ensemble of choice for most of the theoretical calculations of fluctuations, like lattice QCD or effective QCD theories.
    The system created in heavy-ion collisions, however, is different. 
    It is not in contact with the heat bath but rather expands into the vacuum.
    A total conserved charge, like baryon number, is conserved exactly throughout the collision.
    This mismatch between theory and experiment is perhaps the most important conceptual issue, and it reflects the well known fact that the thermodynamics equivalence of statistical ensembles does not extend to event-by-event fluctuations.
    It has been argued that, qualitatively, the conditions mimicking the grand-canonical ensemble can be achieved in heavy-ion collisions by considering fluctuations in a limited acceptance $|y| < 0.5 \, \Delta Y_{\rm accept}$ along the rapidity axis~\cite{Koch:2008ia}, but modifications to this picture are necessary for a quantitative analysis.

    \item Coordinate vs momentum space 
    
    Another important issue lies in the difference between coordinate and momentum spaces.
    Theoretical calculations of fluctuations typically operate in the coordinate space whereas the momentum space is integrated out.
    The opposite situation is encountered in the experiment: the cuts are imposed on the momenta of the particles emerging from a heavy-ion collision, whereas their spatial coordinates are integrated over the entire fireball.
    In some cases, the momenta can be closely correlated to the coordinates and the connection between the two spaces made.
    This happens, for example, in the case of Bjorken flow, where the momentum rapidity $y$ can be identified with the space-time rapidity $\eta_s$ up to the thermal smearing~\cite{Ling:2015yau,Ohnishi:2016bdf}.
    More generally, the transformation between the coordinates and momenta of the particles is typically achieved through the Cooper-Frye formula~\cite{Cooper:1974mv}, although extra care is required for an accurate treatment of fluctuations and correlations.

    \item Proxy observables in the experiment (net-proton, net-kaon) vs actual conserved charges in QCD (net-baryon, net-strangeness)
    
    Because of the experimental limitations concerning the identification of neutral particles, it is usually not possible to measure fluctuations of quantities that are conserved in QCD, with the exception of electric charge.
    Proxy observables thus have to be used, like net proton number instead of the net baryon number or net kaons instead of net strangeness.
    This makes it challenging to analyze the data within theories which are restricted to fluctuations of conserved charges, like lattice QCD.
    On the other hand, in some cases it is possible to reconstruct conserved charge fluctuations from proxy observables, e.g. it is possible to obtain baryon cumulants from the measured proton ones under the assumption of isospin randomization~\cite{Kitazawa:2011wh,Kitazawa:2012at}.

    \item Volume fluctuations
    
    Due to the event-by-event fluctuations in the initial state, which are mainly of a geometrical origin, the fireball volume fluctuates from one event to another.
    These volume fluctuations, which cannot be completely avoided in the experiment, are a source of unwanted non-dynamical fluctuations in the final observables.
    For instance, significant volume fluctuations can make the scaled variance of particle number fluctuations to appear large even in the limit of uncorrelated particle production~\cite{Gorenstein:2011vq}, potentially mistaking this result for a signal of the QCD critical point.
    Controlling volume fluctuations is thus very important, and different strategies are being for this purpose, including correcting either the data or theory calculations for volume fluctuations~\cite{Skokov:2012ds,Luo:2013bmi,HADES:2020wpc}, or constructing observables where the effect of volume fluctuations is minimized~\cite{Gorenstein:2011vq,Sangaline:2015bma,Braun-Munzinger:2016yjz}.

    \item Hadronic phase
    
    The potentially long-lasting hadronic phase  introduces a decorrelation between the numbers of conserved charges at the last point of chemical equilibrium and the final freeze-out~\cite{Steinheimer:2016cir}, thus blurring the potential signals of equilibrium fluctuations.
    Hadronic afterburner codes can be used to estimate these effects.

    \item Non-equilibrium and memory effects
    
    The system created in heavy-ion collisions is a highly dynamical and inhomogeneous one, at best exhibiting equilibrium only locally.
    The fluctuations can thus be notably affected by non-equilibrium effects, in particular the critical slowing down near the critical point may modify even the qualitative features of critical fluctuations~\cite{Mukherjee:2016kyu}.
    Furthermore, as the fluctuations of a conserved quantity in a given acceptance can only emerge through diffusion -- a surface effect -- it is feasible that the measurements would reflect fluctuations averaged over the course of the collision rather than a single point in time.
    On the flip side, this fact has been suggested as a way to probe the QGP phase, which is taking place early in the collision~\cite{Jeon:2000wg,Asakawa:2000wh}.

\end{itemize}

\section{Recent developments}

\subsection{Subensemble acceptance method (SAM)}

The subensemble acceptance method~(SAM) has been developed recently~\cite{Vovchenko:2020tsr,Poberezhnyuk:2020ayn,Vovchenko:2020gne,Vovchenko:2021yen} to perform a correction on the cumulants for exact global conservation laws.
Although this topic has been studied before in the framework of ideal gas of particles and antiparticles~\cite{Begun:2004gs,Bzdak:2012an}, the main feature of the SAM is that it is \emph{model-independent} and has no assumptions about the underlying equation of state.
Thus, it is applicable to various theoretical calculations of conserved charge susceptibilities like lattice QCD.

\subsubsection{Single conserved charge}

The original SAM~\cite{Vovchenko:2020tsr,Poberezhnyuk:2020ayn} was formulated for the case of single conserved charge.
The method explores fluctuations of a globally conserved charge $B$ in a coordinate space subsystem which covers a fraction $\alpha$ of the total volume.
Assuming the thermodynamic limit, $V \to \infty$, as appropriate for central collisions of heavy ions, one can represent the probability $P(B)$ to have a particular value of a conserved charge in the subsystem as a product of canonical partition functions for the subsystem and the complement.
Thermodynamic relations can then be used to relate the cumulants $\kappa_n^B$ of the charge distribution inside the subvolume to the grand-canonical susceptibilities $\chi_n^B$.
In particular, one obtains the following ratios of susceptibilities constrained by exact global conservation:
\begin{align}
\frac{\kappa_2^B}{\kappa_1^B} & = (1-\alpha) \, \frac{\chi_2^B}{\chi_1^B}, \\
\label{eq:skew}
\frac{\kappa_3^B}{\kappa_2^B} & = (1-2\alpha) \, \frac{\chi_3^B}{\chi_2^B}, \\
\label{eq:kurt}
\frac{\kappa_4^B}{\kappa_2^B} & = (1-3\alpha \beta) \, \frac{\chi_4^B}{\chi_2^B} - 3 \alpha \beta \left( \frac{\chi_3^B}{\chi_2^B}\right)^2~.
\end{align}
Here $\beta = 1 - \alpha$.
These expressions indicate that the effect of global conservation disappears in the limit $\alpha \to 0$, as it should, and that for finite values of $\alpha$ the deviations are larger for higher-order cumulants.

\begin{figure}[t]
  \centering
  \includegraphics[width=.49\textwidth]{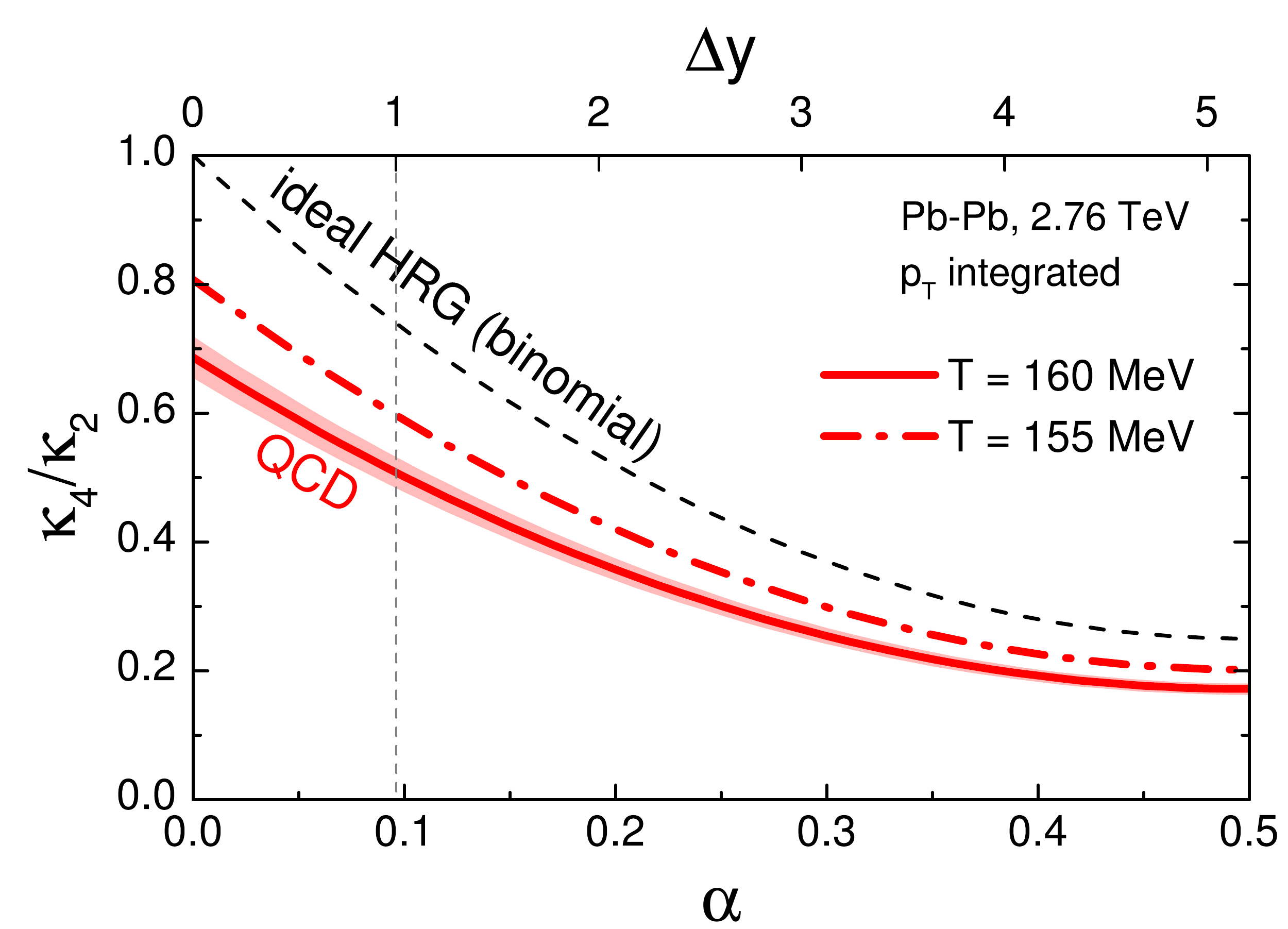}
  \includegraphics[width=.49\textwidth]{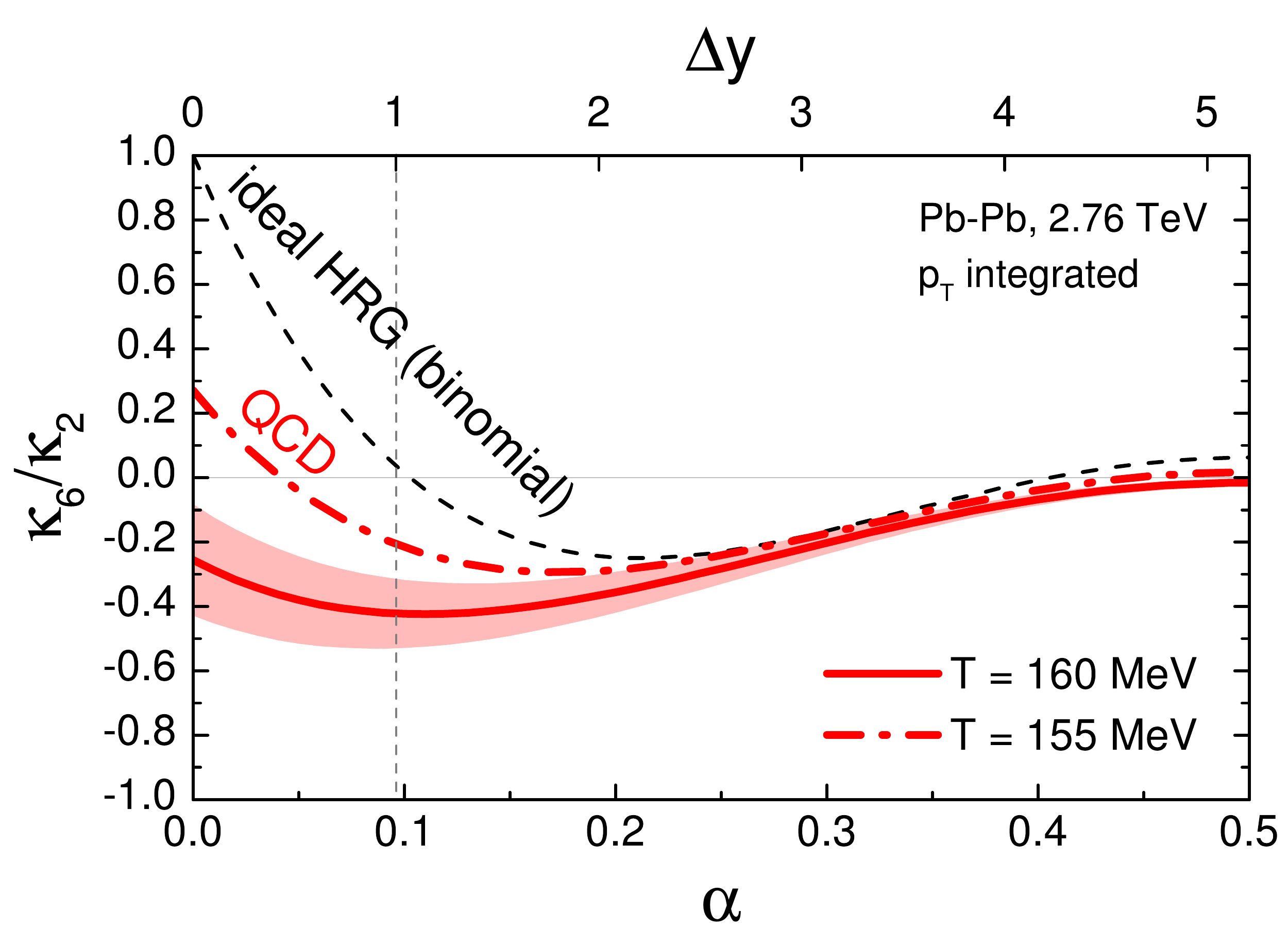}
  \caption{
   Adapted from Ref.~\cite{Vovchenko:2020tsr}.
   The kurtosis $\kappa_4 / \kappa_2$~(left panel) and the hyperkurtosis $\kappa_6 / \kappa_2$~(right panel) of baryon number fluctuations in 2.76 TeV Pb-Pb collisions at the LHC as function of the volume fraction $\alpha$~(rapidity acceptance $\Delta y$) calculated assuming chemical freeze-out at $T \simeq 155-160$~MeV, incorporating exact baryon conservation via the SAM, and using lattice data for baryon number susceptibilities from~\cite{Borsanyi:2018grb}.
  }
  \label{fig:SAMLHC}
\end{figure}

The SAM was used in Ref.~\cite{Vovchenko:2020tsr} in order to calculate the higher-order baryon number fluctuations in central Pb-Pb collisions at the LHC based on first-principle lattice QCD results and incorporating the effect of baryon conservation. These calculations are depicted in Fig.~\ref{fig:SAMLHC} for the kurtosis $\kappa_4 / \kappa_2$ and the hyperkurtosis $\kappa_6 / \kappa_2$ as a function of the acceptance fraction $\alpha$.
This fraction can be related to the acceptance window in rapidity by utilizing the fraction of the accepted charged particles as a proxy, i.e. $\alpha \approx N_{\rm ch}(\Delta y) / N_{\rm ch}(\infty) \approx {\rm erf} \left( \frac{\Delta y}{2 \sqrt{2} \sigma_y} \right)$.
Here $\sigma_y$ is the Gaussian width of the $N_{\rm ch}$ rapidity distribution, which equals $\sigma_y \approx 3.86$ for $\sqrt{s_{\rm NN}} = 2.76$~TeV~\cite{ALICE:2013jfw} and $\sigma_y \approx 4.12$ for $\sqrt{s_{\rm NN}} = 5.02$~TeV~\cite{ALICE:2016fbt}.
The results make strong influence of baryon conservation on baryon fluctuations evident.
Of particular interest is the hyperkurtosis, as the negative value of $\chi_6^B / \chi_2^B$ in the grand-canonical ensemble is thought to be a signal of chiral criticality~\cite{Friman:2011pf}.
To establish this fact experimentally will require measurements of a negative $\kappa_6 / \kappa_2$ in rapidity acceptances $\Delta y \lesssim 1$, as only in this case the corresponding measurements shall be more sensitive to the equation of state than to baryon conservation.
Such measurements are planned in Runs 3 \& 4 at the LHC~\cite{Citron:2018lsq}.

\subsubsection{Multiple conserved charges}

In Ref.~\cite{Vovchenko:2020gne} the SAM has been extended to the case of multiple exactly conserved charges, as appropriate for baryon, electric charge and strangeness in heavy-ion collisions.
In this case one can express the joint $BQS$ cumulants inside the subsystem in terms of the fraction $\alpha$ and the grand-canonical mixed susceptibilities $\chi_{lmn}^{BQS}$.
One can highlight the following observations concerning cumulants of multiple conserved charges:
\begin{itemize}
    \item For all cumulants up to third order, the effect of global conservation laws and equation of state factorizes into a product of cumulants of the Bernoulli and grand-canonical distributions, i.e.
    \begin{equation}
        \kappa_{l,m,n}^{BQS} = \frac{\kappa_{l+m+n}^{\rm Bernoulli}(\alpha)}{\alpha} \times \kappa_{l,m,n}^{BQS,\rm gce}, \quad l+m+n \leq 3.
    \end{equation}
    Here $\kappa_{1}^{\rm Bernoulli}(\alpha) = \alpha$, $\kappa_{2}^{\rm Bernoulli}(\alpha) = \alpha (1-\alpha)$, and $\kappa_{3}^{\rm Bernoulli}(\alpha) = \alpha (1-\alpha) (1-2\alpha)$.
    As a consequence, the conservation effects cancel out in any ratio of two second order cumulants and in any ratio of two third order cumulants, i.e. the corresponding ratios reduce to ratios of the grand canonical susceptibilities, making them suitable for comparisons to theory. This statement extends also to the experimentally more directly accessible correlators of non-conserved quantities with conserved ones, such as proton-charge or kaon-charge correlators.
    
    \item The kurtosis and the higher order cumulants of a conserved charge distribution are affected by conservation laws involving other conserved charges.
    However, the estimates based on the HRG model presented in Ref.~\cite{Vovchenko:2020gne} indicate that, at least for the kurtosis, this effect is small in heavy-ion collisions.
\end{itemize}

\subsubsection{SAM-2.0}

The original SAM allows one to study the conserved charge cumulants measured in the coordinate space subvolume of a uniform thermal system.
Experimental measurements, on the other hand, are performed in the momentum space, and the fireball created in heavy-ion collisions may also be highly inhomogeneous.
Furthermore, the measurements typically deal with non-conserved quantities like (net) protons rather than conserved ones such as (net) baryons.
To address these limitations, an extended formalism -- SAM-2.0 -- was recently formulated in Ref.~\cite{Vovchenko:2021yen}.
The SAM-2.0 allows one to study the effects of exact conservation of charge $B$ on cumulants of a non-conserved quantity $p$ correlated to $B$, in arbitrary acceptance.
The method works as long as the following two assumptions are accurate: (i) the system is sufficiently large such that all the relevant distributions are highly peaked at the mean and (ii) in the absence of exact global conservation the distributions inside the acceptance are uncorrelated to the ones outside the acceptance.
If this is the case, the method allows one to express the joint cumulants of the $p-B$ distribution in the acceptance constrained by the $B$-conservation in terms of cumulants of the unconstrained distribution both inside and outside the acceptance, i.e. it provides the mapping $\tilde{\mathcal{S}}$ such that
\begin{equation}
\kappa_{n,m;\rm in}^{p,B} = \tilde{\mathcal{S}}\left[\kappa_{n,m;\rm in}^{p,B;\rm gce},\kappa_{n,m;\rm out}^{p,B;\rm gce} \right]~.
\end{equation}
For explicit expressions see Ref.~\cite{Vovchenko:2021yen}.

In practice, the SAM-2.0 can be used to correct theoretical calculations of various event-by-event fluctuations, e.g. using relativistic hydrodynamics, for global charge conservation, as it is usually challenging to implement the exact conservation laws in such calculations directly.
The SAM-2.0 was recently used in Ref.~\cite{Vovchenko:2021kxx} to analyze proton number cumulants in central Au-Au collisions at RHIC beam energy scan energies.

\subsection{Subensemble sampler}

The subensemble sampler, recently introduced in Ref.~\cite{Vovchenko:2020kwg}, is a generalization of the Cooper-Frye particlization procedure to the case of interacting hadron resonance gas.
It allows one to incorporate simultaneously the local correlations due to interactions, like the excluded volume effect, and exact global conservation of (multiple) conserved charges.
Being a Monte Carlo sampler, the procedure can be naturally combined with a hadronic afterburner.

\subsubsection{Event-by-event fluctuations at the LHC}

In Ref.~\cite{Vovchenko:2020kwg} the subensemble sampler was used to study various event-by-event fluctuations in 2.76 TeV Pb-Pb collisions at the LHC.
Particlization at $T = 155$~MeV was modeled using a blast-wave hypersurface, followed by a cascade of resonance decays.
The calculation includes the excluded volume effect in the baryonic sector, by utilizing the EV-HRG model of Refs.~\cite{Vovchenko:2016rkn,Vovchenko:2017xad}, and exact conservation of baryon number, electric charge, and strangeness.
The results were compared with experimental data of the ALICE Collaboration, that were measured in the acceptance $0.6 < p < 1.5$~GeV/$c$ and $\Delta \eta_{\rm acc} \leq 1.6$.
The analysis indicates that the variances of net proton and net $\Lambda$ distributions are suppressed by both the excluded volume and baryon conservation, with the latter effect being the more dominant one.
The $\Lambda$ fluctuations are also affected by the strangeness conservation, although this effect is rather small.
The net kaon and net pion fluctuations are notably affected by the strangeness and electric charge conservation, respectively.
Correlations generated by the decays like $\phi \to K^+ K^-$ and $\rho^0 \to \pi^+ \pi^-$ were also found to be important for comparisons with experimental data.

\begin{figure}[t]
  \centering
  \includegraphics[width=.40\textwidth]{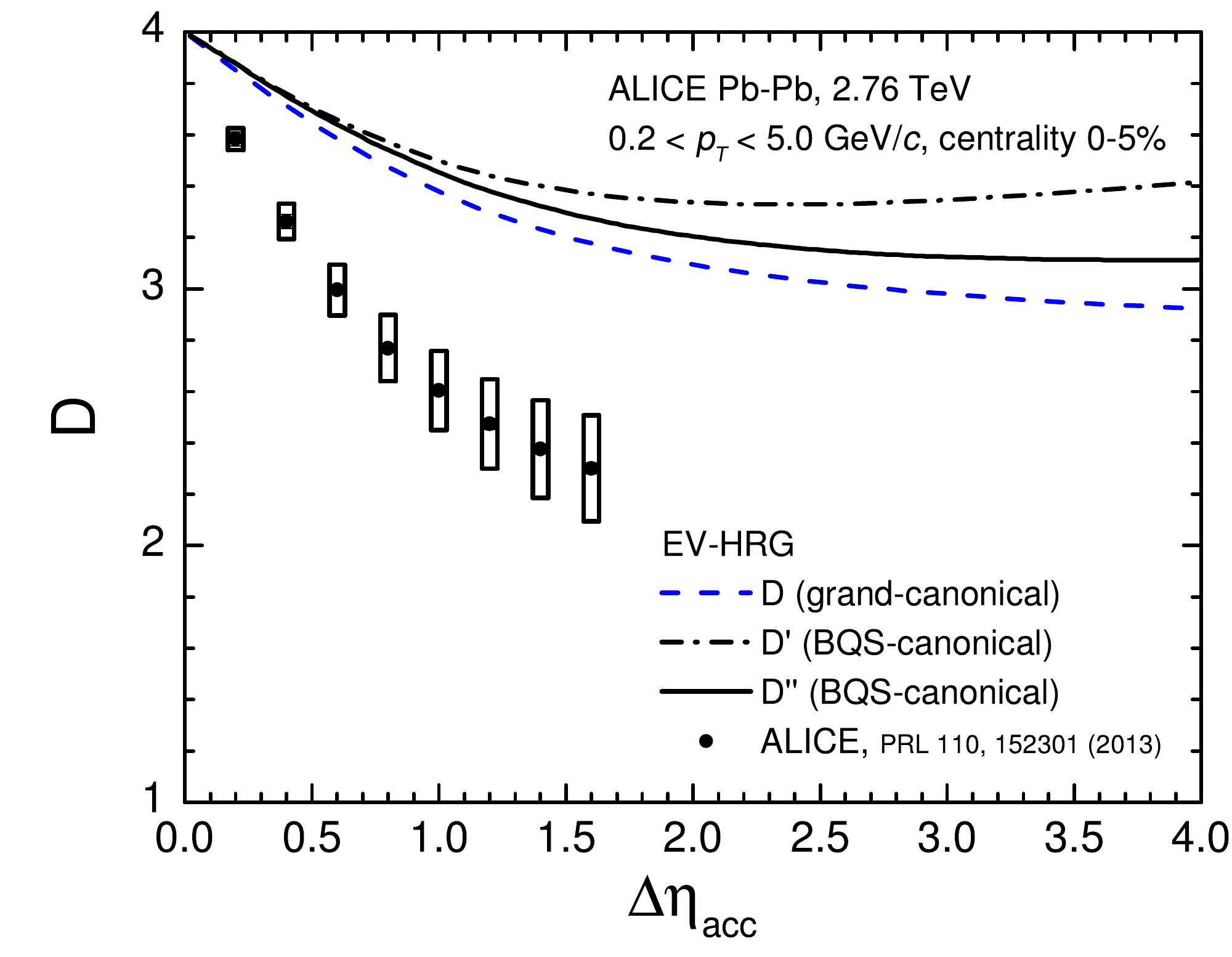}
  \hskip3pt
  \includegraphics[width=.49\textwidth]{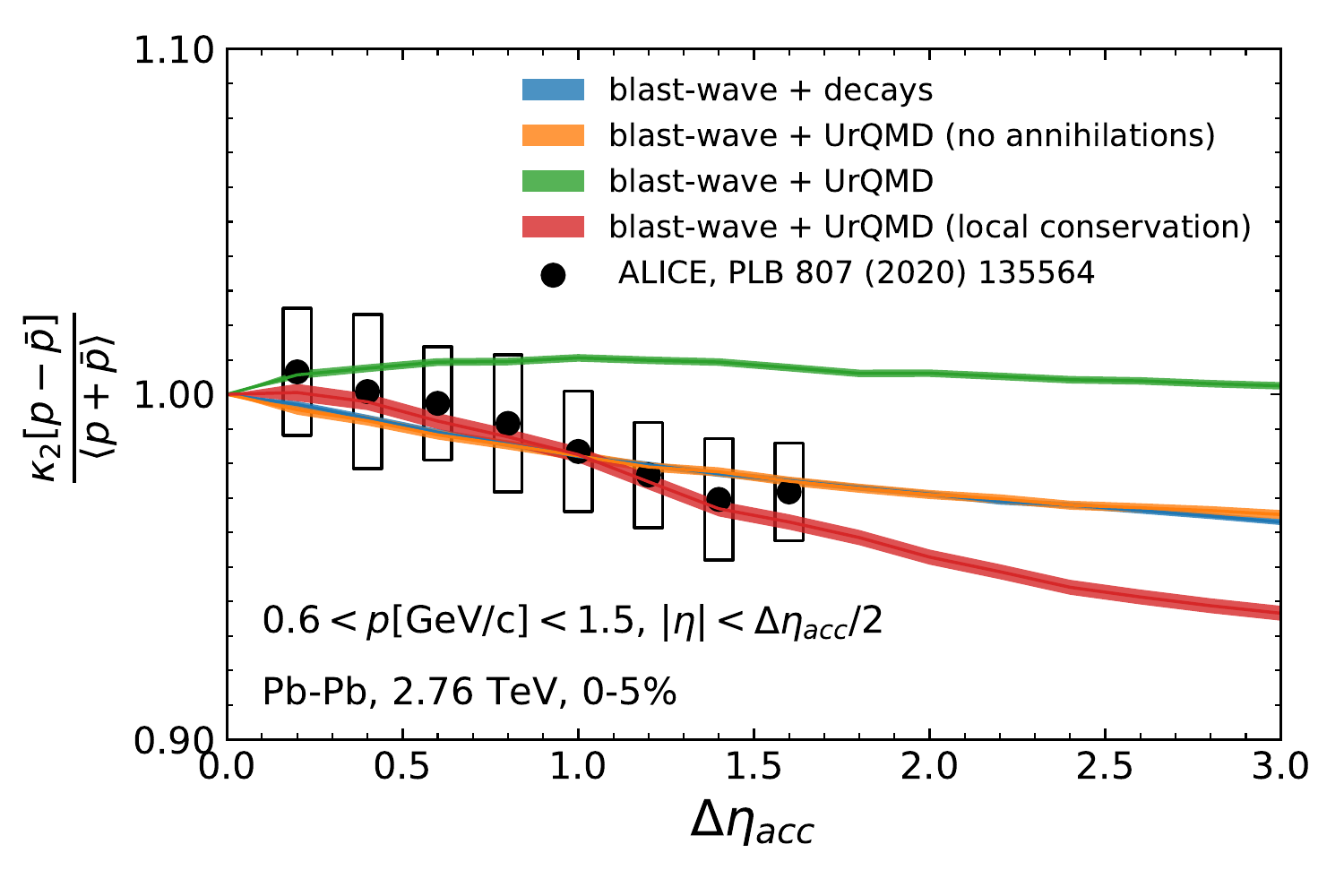}
  \caption{
  Adapted from Refs.~\cite{Vovchenko:2020kwg,Savchuk:2021aog}.
  Pseudorapidity acceptance dependence of the $D$-measure of 
  net charge fluctuations~(\emph{left panel}) and net proton $\kappa_2[p-\bar{p}]/\langle p+\bar{p}\rangle$~(\emph{right panel}) in 0-5\% central 2.76~TeV Pb-Pb collisions at the LHC. The lines and bands depict model calculations with different treatments of conservation laws and baryon annihilation. The experimental data are from Refs.~\cite{ALICE:2012xnj,ALICE:2019nbs}.
  }
  \label{fig:samplerLHC}
\end{figure}

An interesting behavior was obtained for the variance of net charge distribution, namely the so-called $D$-measure, $D = 4 \langle \delta Q^2 \rangle / \langle N_{\rm ch} \rangle$, which was suggested in Ref.~\cite{Jeon:2000wg} as a sensitive probe of QGP.
Generalized measures $D'$~\cite{Pruneau:2002yf} and $D''$~\cite{Bleicher:2000ek} were later introduced to minimize the effect of charge conservation, and these were measured by the ALICE Collaboration in Ref.~\cite{ALICE:2012xnj} in a broad transverse momentum acceptance $0.2 < p_T < 5.0$~GeV/$c$ as function of $\Delta \eta_{\rm acc}$.
The comparison between the data and the subensemble sampler based calculations is depicted in the left panel of Fig.~\ref{fig:samplerLHC}, indicating that the model fails to quantitatively describe the suppression of the $D$-measure seen in the experiment.
One tantalizing possible explanation here is the QGP formation, where the suppression of the $D$-measure would be expected~\cite{Jeon:2000wg}, but more studies are required to reach a firm conclusion.

\subsubsection{Constraining baryon annihilation with fluctuations}

In Ref.~\cite{Savchuk:2021aog} it was pointed out how event-by-event fluctuations can used to constrain the effect of baryon annihilation in the hadronic phase of heavy-ion collisions, which is known to be challenging to model properly in transport codes due to the multi-particle backreaction.
Namely, the variance of net-baryon distribution normalized by the Skellam distribution baseline, $\kappa_2[B-\bar{B}]/\langle B+\bar{B}\rangle$, is sensitive to the possible modification of (anti)baryon yields due to $B\bar{B}$ annihilation.
This can be explained as follows: to leading order the annihilation leaves the variance $\kappa_2[B-\bar{B}]$ of net baryon number in a sufficiently large acceptance unchanged, but the mean total baryon yield $\langle B+\bar{B}\rangle$ can be significantly modified, leading to the enhancement of $\kappa_2[B-\bar{B}]/\langle B+\bar{B}\rangle$ if the baryon yields are suppressed. This enhancement can be expected to survive in the experimentally accessible fluctuations of net proton number.
This is illustrated in a model calculation in the right panel of Fig.~\ref{fig:samplerLHC} by utilizing the subensemble sampler and the  hadronic afterburner UrQMD with and without the $B\bar{B}$ annihilation reactions.
The available experimental data of the ALICE
Collaboration on net proton fluctuations~\cite{ALICE:2019nbs} disfavors a notable suppression of (anti)baryon yields due to baryon annihilation predicted by the present version of UrQMD if only global baryon conservation is incorporated.
On the other hand, the annihilations were shown to improve the data description when local baryon conservation with rapidity range $\Delta y_{\rm cons} = 3$ is imposed instead of global baryon conservation.
The two effects can be disentangled by measuring $\kappa_2[B+\bar{B}]/\langle B+\bar{B}\rangle$,
which at the LHC was shown to be notably suppressed by annihilations but virtually unaffected by baryon number conservation.

\subsection{Proton number cumulants from hydrodynamics}

Fluctuations of (net-)proton number are one of the primary observables in RHIC-BES used to probe the QCD critical point on the phase diagram.
Making quantitative predictions for the critical point signatures in heavy-ion collisions is challenging and currently much work is being done on that avenue within the framework of hydrodynamics with critical fluctuations~\cite{Stephanov:2017ghc,Bluhm:2020mpc,An:2021wof}.
The issue can also be approached from a different angle: the critical point signal, if it exists, should manifest itself in deviations from non-critical baseline expectations that do not incorporate any critical point effects. The baseline can be obtained through realistic dynamical modeling of heavy-ion collisions.

This task has recently been achieved in Ref.~\cite{Vovchenko:2021kxx} in the framework of (3+1)D relativistic viscous hydrodynamics simulations of 0-5\% central Au-Au collisions at RHIC-BES energies. 
The calculations utilize collision geometry based 3D initial conditions~\cite{Shen:2020jwv}, a lattice-based crossover-type equation of state NEOS-BQS~\cite{Monnai:2019hkn}, and a hydrodynamics solver MUSIC~\cite{Schenke:2010nt}.
The proton cumulants are computed analytically at Cooper-Frye particlization, incorporating the essential non-critical contributions to proton number cumulants such as global baryon conservation and short-range repulsive interactions between baryons, modeled through the excluded volume effect.

\begin{figure}[t]
  \centering
  \includegraphics[width=.32\textwidth]{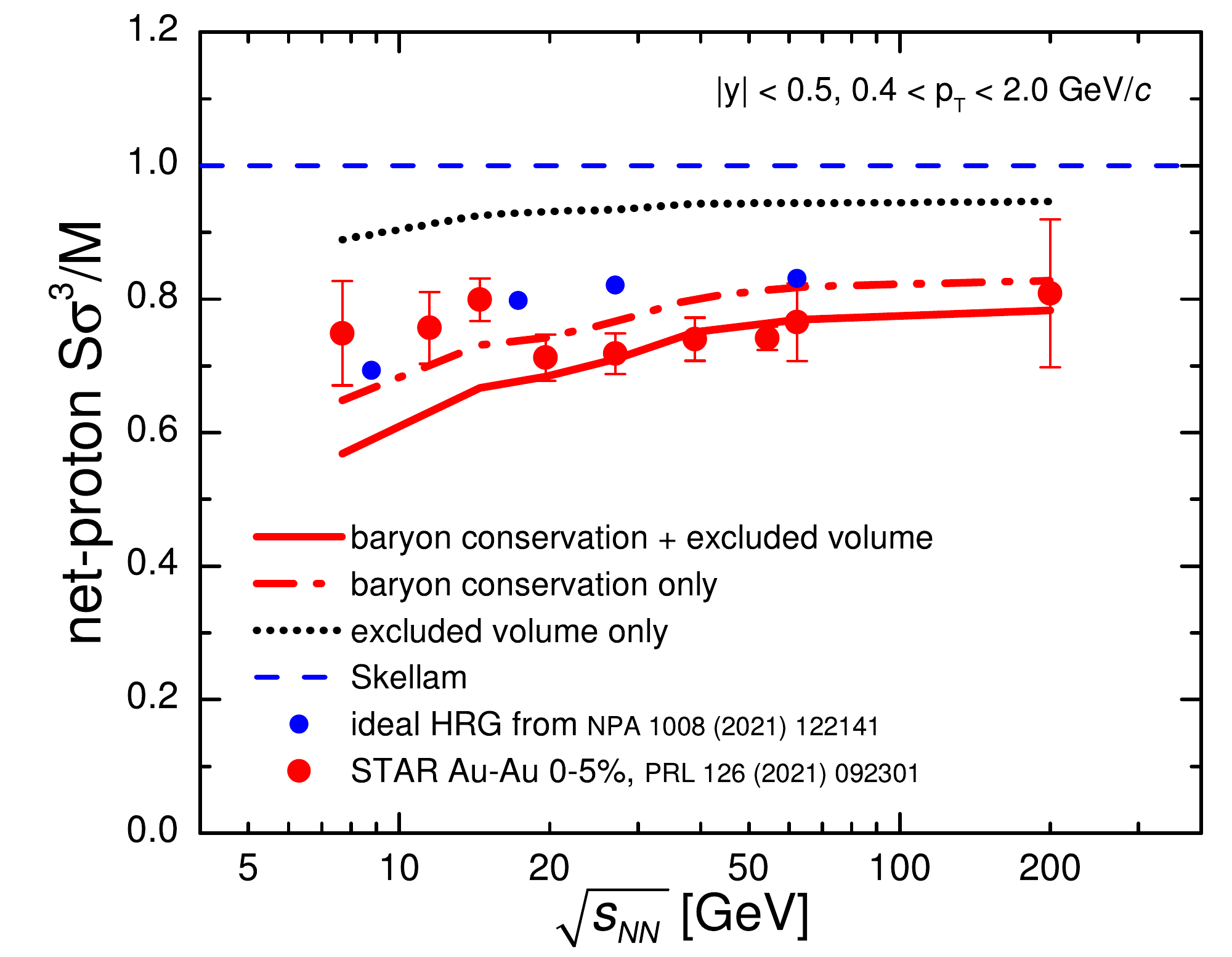}
  \includegraphics[width=.32\textwidth]{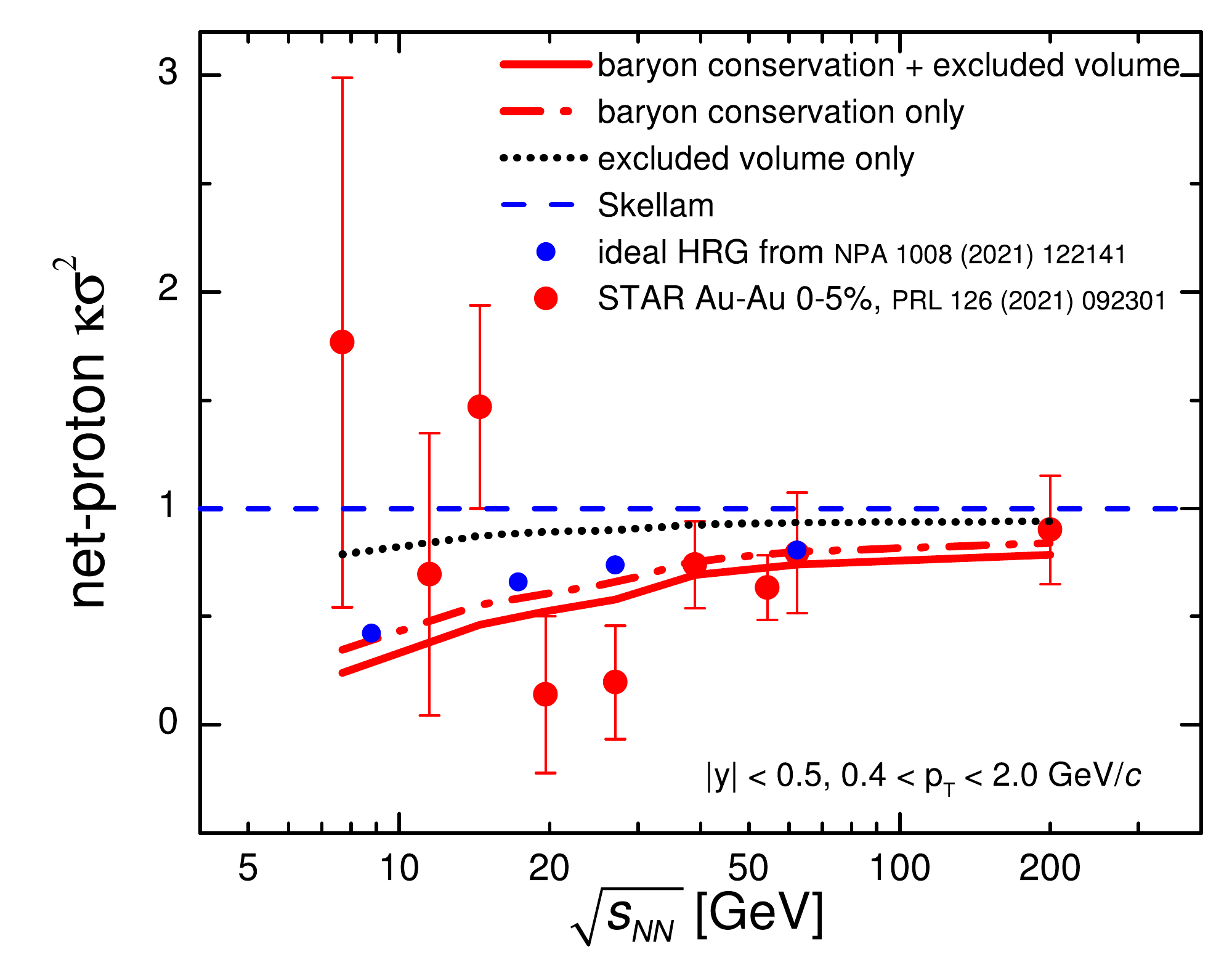}
  \includegraphics[width=.32\textwidth]{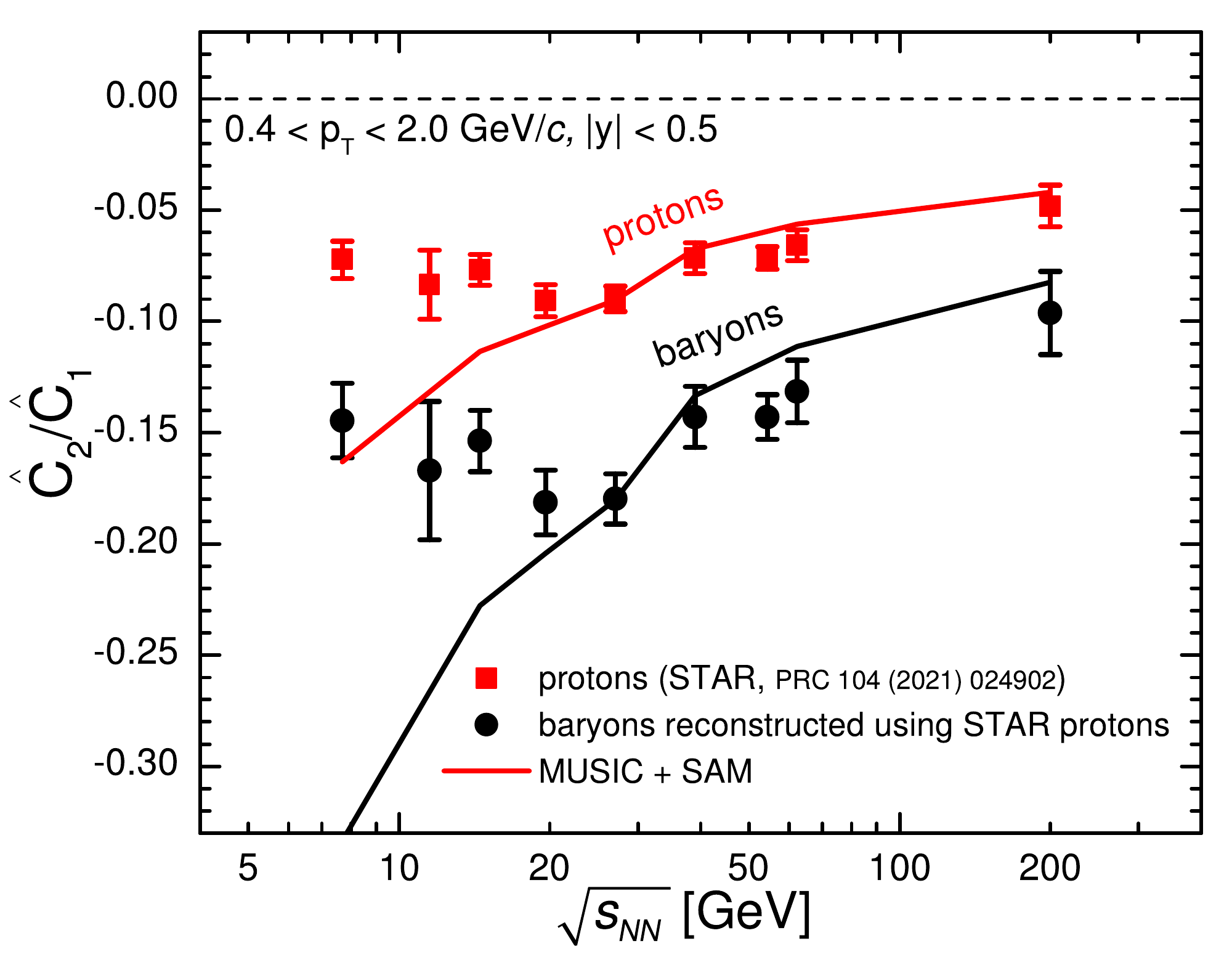}
  \caption{
  From Ref.~\cite{Vovchenko:2021kxx}.
 Collision energy dependence of the net proton cumulant ratios $\kappa_3/\kappa_1$~(\emph{left panel}) and $\kappa_4/\kappa_2$~(\emph{middle panel}), and the scaled second factorial cumulant $\hat{C}_2/\hat{C}_1$ of protons and baryons~(\emph{right panel}). 
 The experimental data are from Refs.~\cite{STAR:2020tga,STAR:2021iop}.
  }
  \label{fig:RHIC}
\end{figure}

The left and middle panels of Fig.~\ref{fig:RHIC} show the $\sqrt{s_{\rm NN}}$ dependence of the net proton cumulant ratios $\kappa_3/\kappa_1 \equiv S \sigma^3 / M$ and $\kappa_4/\kappa_2 \equiv \kappa \sigma^2$.
These ratios are suppressed relative to unity by both the baryon conservation and the excluded volume, with the former effect having being stronger and its magnitude in line with the earlier findings of Ref.~\cite{Braun-Munzinger:2020jbk}.
Comparisons with the experimental data of the STAR Collaboration~\cite{STAR:2020tga} indicate that both the baryon conservation and repulsion are needed to describe the data quantitatively at $\sqrt{s_{\rm NN}} \gtrsim 20$~GeV while at lower energies the data for $S \sigma^3 / M$  indicate a smaller suppression than predicted.

Complementary to the ordinary cumulants, the (anti)proton correlation functions~(factorial cumulants) $\hat{C}_k$ have been suggested as sensitive probes of the genuine multi-particle correlations expected near the critical point~\cite{Bzdak:2016sxg}.
The calculations within the non-critical setup indicate only mild multi-proton correlations which are refleced in small magnitude of the ratios $\hat{C}_3/\hat{C}_1$ and $\hat{C}_4/\hat{C}_1$, which is in agreement with the available experimental data~\cite{STAR:2021iop}. 
It should be noted, though, that error bars are large at $\sqrt{s_{\rm NN}} \lesssim 20$~GeV and the presence of sizable multi-proton correlations in the data for those energies cannot be ruled out.
The second factorial cumulants of both the protons and antiprotons indicate negative two-particle correlations at all the studied collision energies~(Fig.~\ref{fig:RHIC}, right panel). The calculation results for protons agree with the experimental data at $\sqrt{s_{\rm NN}} \geq 20$~GeV but overestimate the strength of negative correlations at lower collision energies.
One tantalizing possibility to explain this disagreement is presence of sizable attractive interactions among protons due to approaching the critical point, although other potential explanations like volume fluctuations should also be considered.

Another important issue is the difference between the cumulants of protons and baryons, as the former ones are measured in the experiment but only the latter ones are accessible in most theories.
This difference is shown in the right panel of Fig.~\ref{fig:RHIC}, illustrating that the measurement of only protons from all the baryons essentially amounts to an additional detector efficiency.
It has been argued that baryon cumulants can be reconstructed from the measured proton ones via a binomial unfolding~\cite{Kitazawa:2012at}, and the result of such a procedure applied to the STAR data is shown by the black symbols. Indeed, once the unfolding has been performed, the agreement between the model and the data for the scaled second factorial cumulant of baryons at $\sqrt{s_{\rm NN}} \gtrsim 20$~GeV is recovered. With the future high-statistics data from BES-II it should be also possible to perform such an unfolding for the high-order cumulants as well.

\section{Summary}

Event-by-event fluctuations are a powerful tool to explore the QCD phase diagram with heavy-ion collisions. A lot of care is required when comparing theory and experiment, however.
The recent advances have clarified important roles played by exact conservation laws, thermal smearing, and repulsive interactions, and provided tools to deal with them.
This will play important role in the future efforts to decode the QCD phase structure from fluctuation measurements in the running and future heavy-ion experiments.

\section*{Acknowledgements}

The author thanks Mark Gorenstein, Volker Koch, Roman Poberezhnyuk, Anar Rustamov, Oleh Savchuk, Chun Shen, Jan Steinheimer, Horst Stoecker, and Nu Xu for fruitful discussions. This work was supported through the
Feodor Lynen Program of the Alexander von Humboldt
foundation, the U.S. Department of Energy, 
Office of Science, Office of Nuclear Physics, under contract number 
DE-AC02-05CH11231231, and within the framework of the
Beam Energy Scan Theory (BEST) Topical Collaboration.

\setlength{\bibsep}{0pt plus 0.4ex}

\end{document}